\documentclass{PoS}

\def\beq{\begin{equation}}
\def\eeq{\end{equation}}
\def\bea{\begin{eqnarray}}
\def\eea{\end{eqnarray}}
\def\beqn{\begin{eqnarray}} \def\eeqn{\end{eqnarray}}

\def\nn{\nonumber}

\def\ln#1{\mathrm{ln}\left(#1\right)}

\newcommand\as{a_{\mathrm{S}}}
\newcommand\alphas{\alpha_{\mathrm{S}}}

\def\beq{\begin{equation}} \def\eeq{\end{equation}}
\def\beqn{\begin{eqnarray}} \def\eeqn{\end{eqnarray}}
 \def\to{\rightarrow}
\def\nn{\nonumber}

\title{
\vspace*{-2.5cm}
Higher-order QED effects in hadronic processes}

\ShortTitle{Higher-order QED effects in hadronic processes}

\author{\speaker{Germ\'an F. R. Sborlini}$^{\ a,b}$\\\\
        $^a$Dipartimento di Fisica, Universit\`a di Milano and INFN Sezione di Milano,
I-20133 Milan, Italy.\\\
        $^b$Instituto de F\'{\i}sica Corpuscular, Universitat de Val\`{e}ncia -- 
Consejo Superior de Investigaciones Cient\'{\i}ficas, Parc Cient\'{\i}fic, E-46980 Paterna, Valencia, Spain.\\\\
        E-mail: \email{german.sborlini@unimi.it}}

\abstract{In this presentation, we describe the computation of higher-order QED effects relevant in hadronic collisions. In particular, we discuss the calculation of mixed QCD-QED one-loop contributions to the Altarelli-Parisi splittings functions, as well as the pure two-loop QED corrections. We explain how to extend the DGLAP equations to deal with new parton distributions, emphasizing the consequences of the novel corrections in the determination (and evolution) of the photon distributions.}

\FullConference{EPS-HEP 2017, European Physical Society conference on High Energy Physics\\
		5-12 July 2017\\
		Venice, Italy}

\begin{document}

\section{Introduction}
\label{sec:Intro}
Due to the improved accuracy and precision of the experimental measurements, the corresponding theoretical calculations must start including effects previously neglected. In the context of hadron colliders, it is a well-known fact that QCD is the dominating interaction, and in consequence the most important corrections are related to the strong force. However, since ${\cal O}(\alphas^2)\approx {\cal O}(\alpha)$ for typical collision energies at the LHC, it becomes necessary to take into account also electroweak (EW) corrections. This is one of the reasons behind the recent explosion in the number of processes that have been computed including EW corrections beyond the leading-order (LO).

The idea of this article is to summarize the recent developments carried out by our group. We start by presenting, in Sec. \ref{sec:DGLAPsplittings}, the extension of the DGLAP equations \cite{Altarelli:1977zs} to include QED effects, as well as a description of the new PDFs associated to leptons and photons. After that, we center in the calculation of the evolution kernels of these equations, namely the splitting functions. In Sec. \ref{sec:Abelianization}, we present an algorithm that allowed us to compute one-loop mixed QCD-QED and two-loop QED corrections to the splitting functions by properly transforming the well-known NLO QCD expressions available in the literature \cite{Curci:1980uw,Furmanski:1980cm,Ellis:1996nn}. Then, we motivate a generalization of the Abelianization algorithm and apply it to a physical process. The selected process is diphoton production; in Sec. \ref{sec:diphoton} we describe the computation of the fully consistent NLO QED corrections obtained through the Abelianization of the \texttt{2gNNLO} code \cite{Catani:2011qz}. Finally, we present the conclusions and briefly discuss some open questions in Sec. \ref{sec:conclusions}.


\section{Extended DGLAP equations and splitting functions}
\label{sec:DGLAPsplittings}
The Altarelli-Parisi equations were originally formulated to describe the perturbative evolution of parton distribution functions (PDF) in the context of QCD interactions \cite{Altarelli:1977zs}. Our purpose is to deal with QCD partons (i.e. gluons and quarks) as well as photons. Moreover, since photons couple to charged leptons, we must also consider their presence inside hadrons and define the associated PDFs. Explicitly, given the canonical basis of PDFs, ${\cal B}_c = \{q_i,l_i,\bar{q}_i,\bar{l}_i,g,\gamma \}$, the extended DGLAP equations read
\beqn
\frac{dF_i}{dt}= \sum_{f} P_{F_i f} \otimes f + \sum_{f} P_{F_i \bar{f}} \otimes \bar{f} +  P_{F_i g} \otimes g +  P_{F_i \gamma} \otimes \gamma \, ,
\label{eq:evolucionFermiones}
\\ \frac{dg}{dt}= \sum_{f} P_{g f} \otimes  f + \sum_{f} P_{g \bar{f}} \otimes  \bar{f} +  P_{g g} \otimes g +  P_{g \gamma} \otimes \gamma \, ,
\label{eq:evolucionGluon}
\\ \frac{d\gamma}{dt}= \sum_{f} P_{\gamma f} \otimes  f + \sum_{f} P_{\gamma \bar{f}} \otimes  \bar{f} +  P_{\gamma g} \otimes g +  P_{\gamma \gamma} \otimes \gamma \, ,
\label{eq:evolucionPhoton}
\eeqn
where $t=\ln{\mu^2}$ is the evolution variable (with $\mu$ the factorization scale), $\otimes$ denotes the convolution operator and $P_{ij}$ are the extended splitting functions. The sum over fermions $f$ runs over all the active flavours of quarks ($n_F$) and leptons ($n_L$).

In order to simplify the previous equations, it is convenient to change the PDF basis as suggested in Ref. \cite{Roth:2004ti}. While the equations for the photon and gluon distributions remains the same, we find
\beqn
\nn && \frac{dF_{v_i}}{dt} =   P_{F_i}^-     \otimes F_{v_i}  +\sum_{j=1}^{n_F} \Delta P_{F_i q_j}^S  \otimes  q_{v_j} + \Delta P_{F_i l}^S \otimes \left(\sum_{j=1}^{n_L}   l_{v_j} \right)    \, ,\quad \frac{d \Delta_{2}^l }{dt} =  P_{l}^+ \otimes \Delta_{2}^l \, ,
\\  && \frac{d \{ \Delta_{uc} , \Delta_{ct} \}}{dt} =  P_{u}^+ \otimes \{ \Delta_{uc} , \Delta_{ct}  \} \, , \quad \frac{d \{ \Delta_{ds} ,\Delta_{sb}  \}}{dt} = P_{d}^+ \otimes \{ \Delta_{ds} ,\Delta_{sb}  \} \, , 
\label{eq:DGLAP1}
\eeqn
for the valence ($F_v=F_v-\bar{F}_v$ with $F$ any fermion) and $\Delta$ distributions. The evolution equations for the other elements of the optimized basis are a bit cumbersome, and they can be found in Refs. \cite{deFlorian:2015ujt,deFlorian:2016gvk,Sborlini:2016dfn}. It is worth appreciating that further simplifications take place when analyzing the ${\cal{O}}(\alpha \alphas)$ and ${\cal{O}}(\alpha^2)$ contributions, since many of the modified splitting kernels are trivially vanishing. 

On the other hand, the sum rules are also extended to include the presence of QED interactions. As usual, we have to impose the conservation of the fermion number inside the proton, as well as the conservation of the total momenta, which must be carried by each possible constituent. Explicitly, this means that the splitting kernels must fulfill
\beq
\int_0^1 \, dx P^-_f =0 \, , \quad \int_0^1 \, dx \, x \, \left(\frac{dg}{dt}+\frac{d\gamma}{dt}+\sum_{f}\frac{df}{dt}\right) \, , 
\eeq
where we are summing over all the possible flavours of fermions and anti-fermions. These conditions fix the behaviour of the regularized splitting kernels in the end-point region, i.e. $x=1$.

To conclude this section, let's make some general remarks. The most prominent consequence of introducing QED interactions is the need to take into account a photon PDF. Moreover, the presence of QED interactions introduces charge separation effects, which were absent in the pure QCD model. Thus, the evolution of PDFs associated to different flavours might differ and this could have a phenomenological impact. In the same direction, the extended QCD-QED model includes quark-lepton mixing, although it is expected to be highly suppressed.


\section{Recovering QED corrections: Abelianization algorithm and splitting functions}
\label{sec:Abelianization}
An interesting property of EW corrections is that we can exploit the previous knowledge of QCD calculations to partially recover them. Explicitly, it is possible to obtain QED and mixed QCD-QED contributions by replacing gluons with photons: this is what we called the \emph{Abelianization} technique. In Refs. \cite{deFlorian:2015ujt,deFlorian:2016gvk} we apply this idea to compute ${\cal O}(\alpha \alphas)$ and ${\cal O}(\alpha^2)$ corrections to the Altarelli-Parisi kernels. Some of these contributions have been computed in Refs. \cite{SPLITTINGS,Sborlini:2015jda}, centering in the amplitude-level results and the multiple collinear behaviour with photon emissions. 

In the context of mixed QCD-QED corrections, the first step consisted in proposing a complete perturbative expansion of splitting kernels in both couplings, i.e.
\beq
P_{ij} = \as \, P_{ij}^{(1,0)} + a \, P_{ij}^{(0,1)} + \as^2 \, P_{ij}^{(2,0)} +  \as a \, P_{ij}^{(1,1)} +  a^2 \, P_{ij}^{(0,2)} + \ldots \, , 
\eeq
and calculate $P_{ij}^{(1,1)}$ and $P_{ij}^{(0,2)}$ by considering $P_{ij}^{(2,0)}$ and replacing one and two gluons by photons, respectively. Of course, this replacement involves some subtleties \cite{deFlorian:2015ujt}. For instance, the Abelianization of $P^{(2,0)}_{gg}$ leads to $P^{(1,1)}_{g\gamma}$ and $P^{(1,1)}_{\gamma g}$; there are two possibilities for replacing one gluon and two different diagrams are obtained. However, starting from $P^{(2,0)}_{qq}$ and implementing the mentioned replacement, we end up with two diagrams contributing only to $P^{(1,1)}_{qq}$. 

\begin{figure}[htb]
\begin{center}
\includegraphics[width=0.44\textwidth]{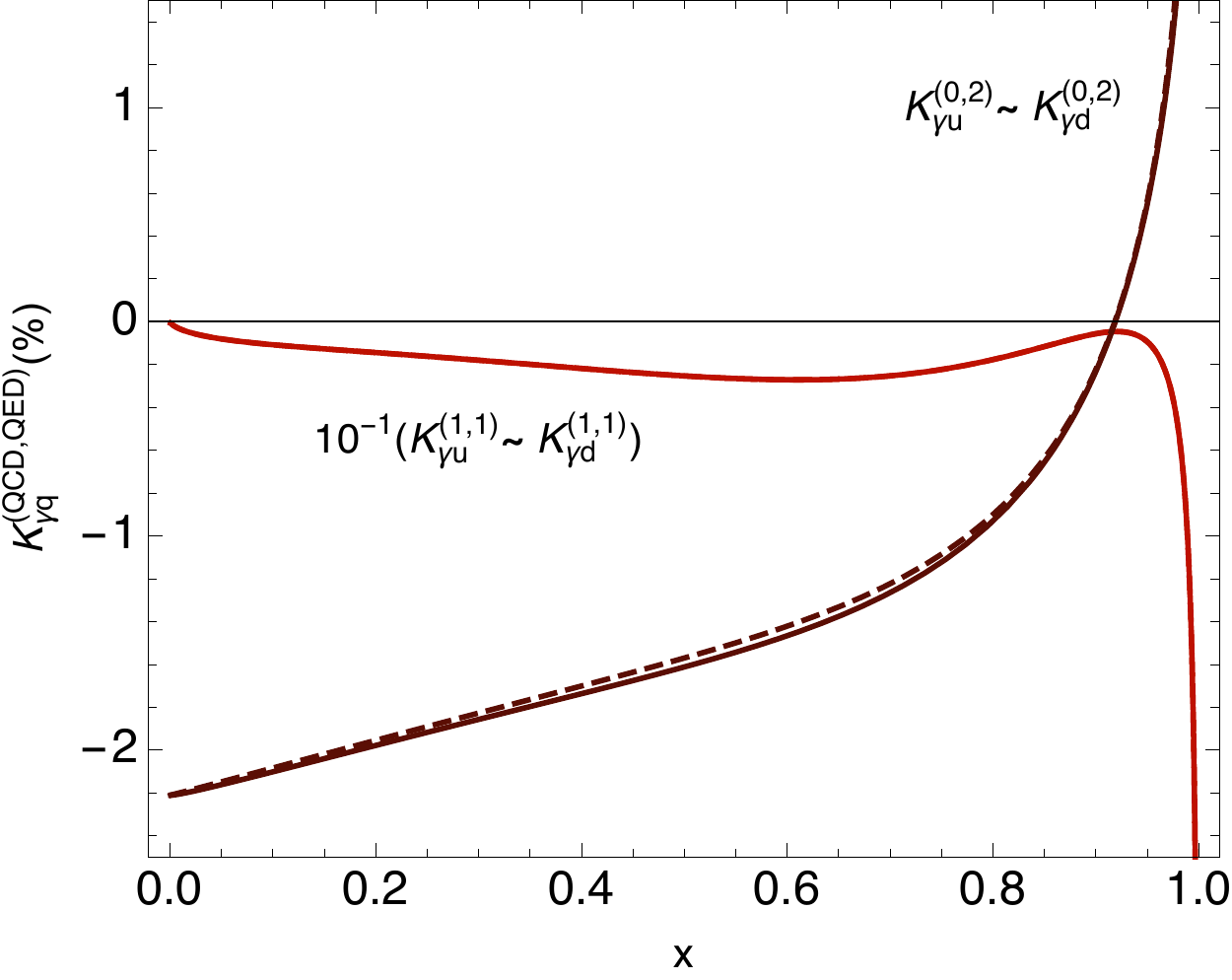} $\quad$
\includegraphics[width=0.44\textwidth]{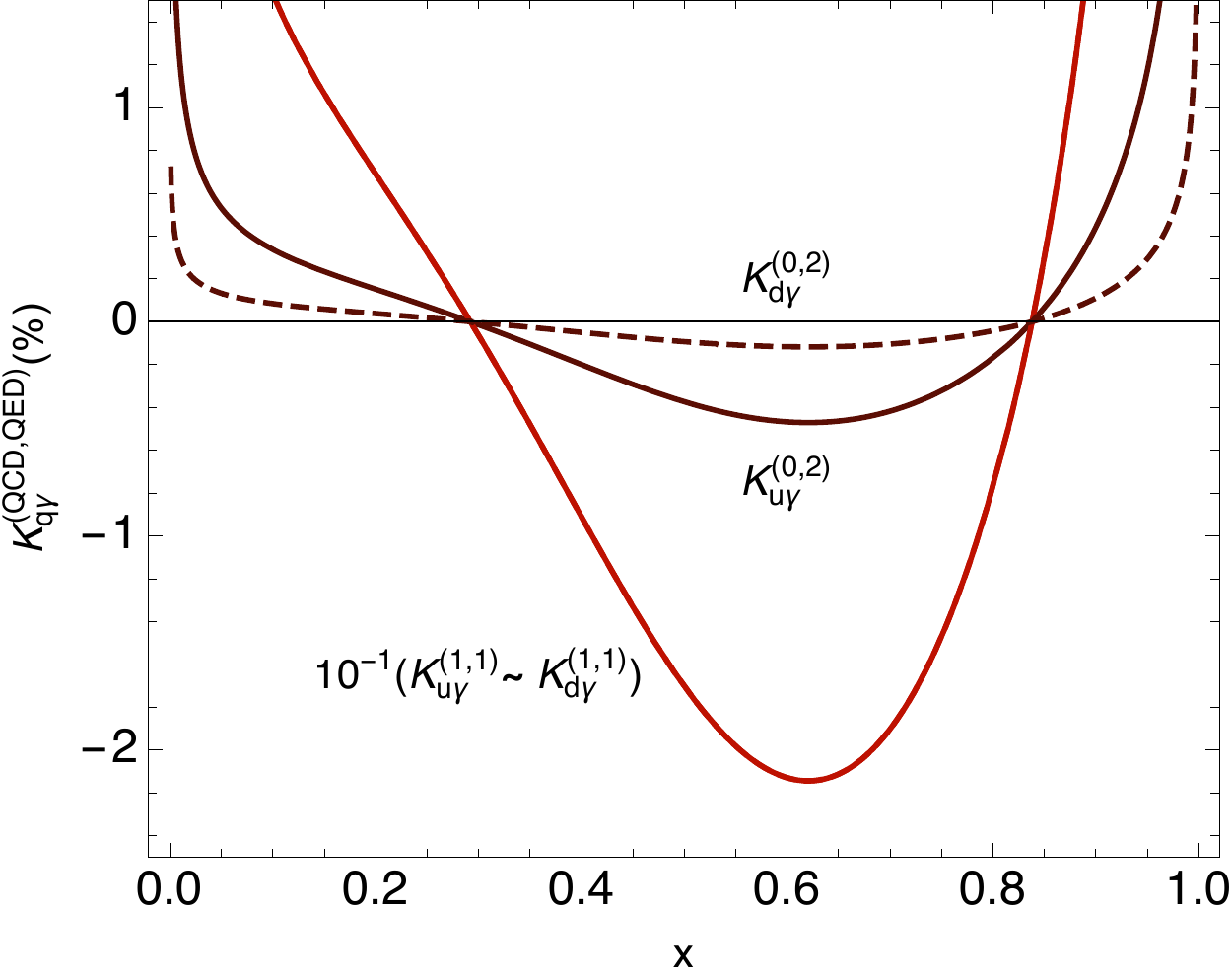} 
\caption{$K$ factors associated to QED corrections to the $P_{\gamma q}$ (left) and $P_{q \gamma}$ (right) splitting functions. The red lines denote the ${\cal O}(\alpha \alphas)$ contributions, whilst the brown ones are ${\cal O}(\alpha^2)$ terms. We can appreciate that the two-loop QED corrections are strongly suppressed in comparison to the mixed QCD-QED ones (factor $10$).}
\label{fig:Splittings}
\end{center}
\end{figure} 

Another subtlety is related to the treatment of the factor $n_F$. In the context of QCD, it is related to the presence of quark loops. However, when we replace gluons by photons, we distinguish the particles according to their electric charge. Moreover, we could include also leptons inside the loop. So, we have the replacement
\beq
n_F \to \sum_f \, e_f ^2 \, , 
\eeq
where the sum is restricted to quarks at ${\cal O}(\alpha \alphas)$, but all fermions are allowed at ${\cal O}(\alpha^2)$. In the last case, it is necessary to explicitly indicate the color degeneration, i.e. we need to sum over each possible quark color. More details about the implementation of the Abelianization algorithm and the replacements implemented can be found in Refs. \cite{deFlorian:2015ujt,deFlorian:2016gvk}. 

Finally, after obtaining the $P^{(1,1)}_{ab}$ and $P^{(0,2)}_{ab}$ terms in the perturbative expansion, we define the ratio $K_{ab}^{(i,j)}=\as^i \, a^j \, P_{ab}^{(i,j)}/P_{ab}^{\rm LO}$ with the \emph{leading-order} kernel $P_{ab}^{\rm LO}=\as \, P_{ab}^{(1,0)}+ a \, P_{ab}^{(0,1)}$. Some illustrative plots are shown in Fig. \ref{fig:Splittings}. In particular, we consider the corrections to $P_{\gamma q}$ (left panel) and $P_{q \gamma}$ (right panel), both at ${\cal O}(\alpha \alphas)$ and ${\cal O}(\alpha^2)$. For splittings involving at least one photon, the lowest order is $P_{ab}^{\rm LO}= a \, P_{ab}^{(0,1)}$ and mixed QCD-QED corrections are dominant when compared with two-loop QED terms (by a factor $10$). The charge separation effect becomes more noticeable in $P_{q \gamma}$, although it is still present in $P_{\gamma q}$.


\section{Benchmark example: diphoton production}
\label{sec:diphoton}
Finally, let's present a practical application of the Abelianization algorithm to a complete physical process. We consider diphoton production in hadron colliders, and we rely on the code \texttt{2gNNLO} \cite{Catani:2011qz}. This code makes use of the $q_T$-subtraction method \cite{Catani:2007vq} to implement a fully differential cross-section calculation including NNLO QCD corrections.

As a proof of concept, we focus in the NLO QCD part of the code and apply the Abelianization technique to obtain the corresponding NLO QED corrections. We transformed both the hard coefficients (and their finite contributions), as well as the universal coefficients used to build the counter-terms. We check the consistency of the this approach by studying the collinear limits and comparing the behaviour of the counter-terms with the previously known QED splitting functions.

\begin{figure}[htb]
\begin{center}
\includegraphics[width=0.49\textwidth]{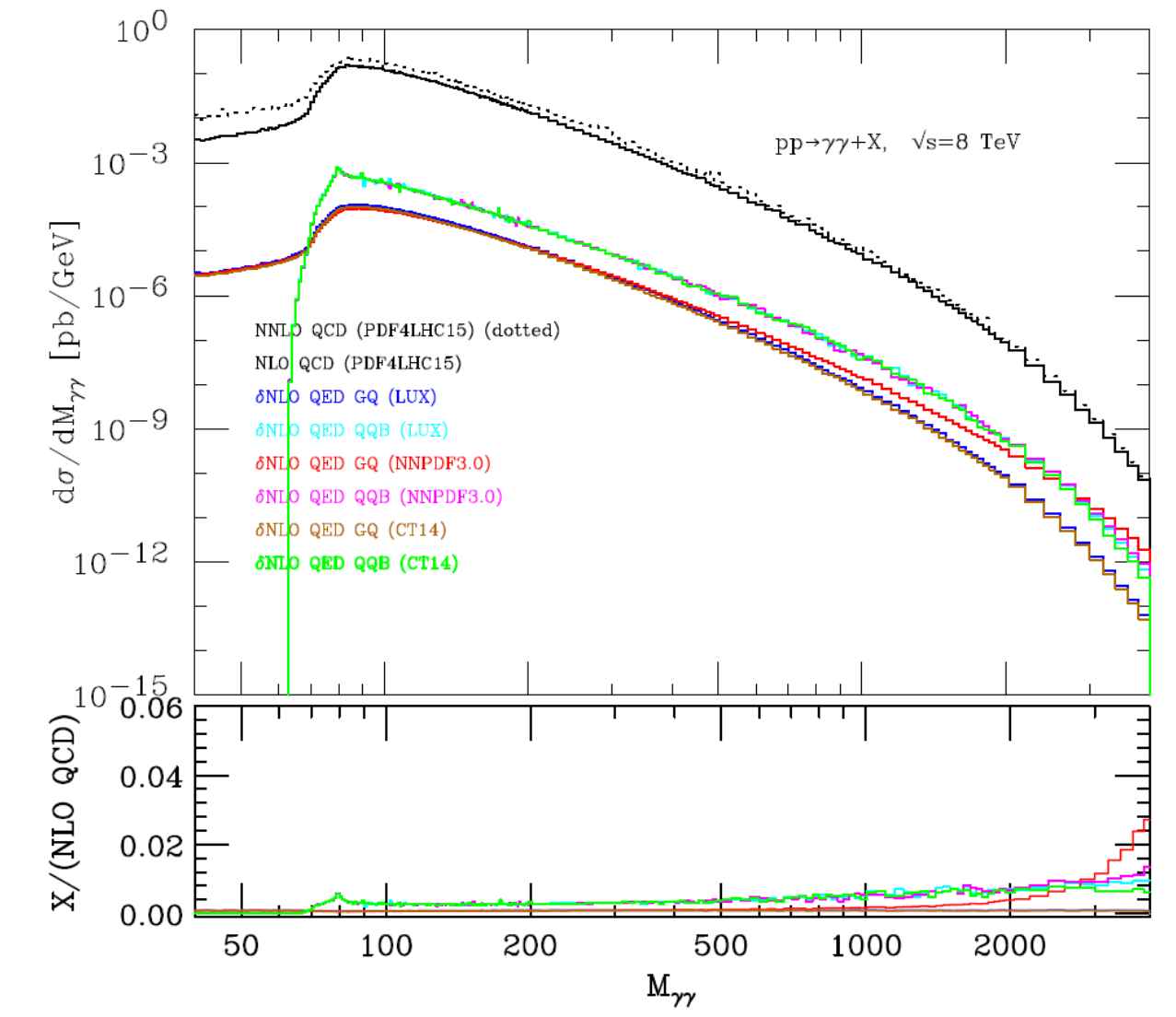} $\quad$
\includegraphics[width=0.47\textwidth]{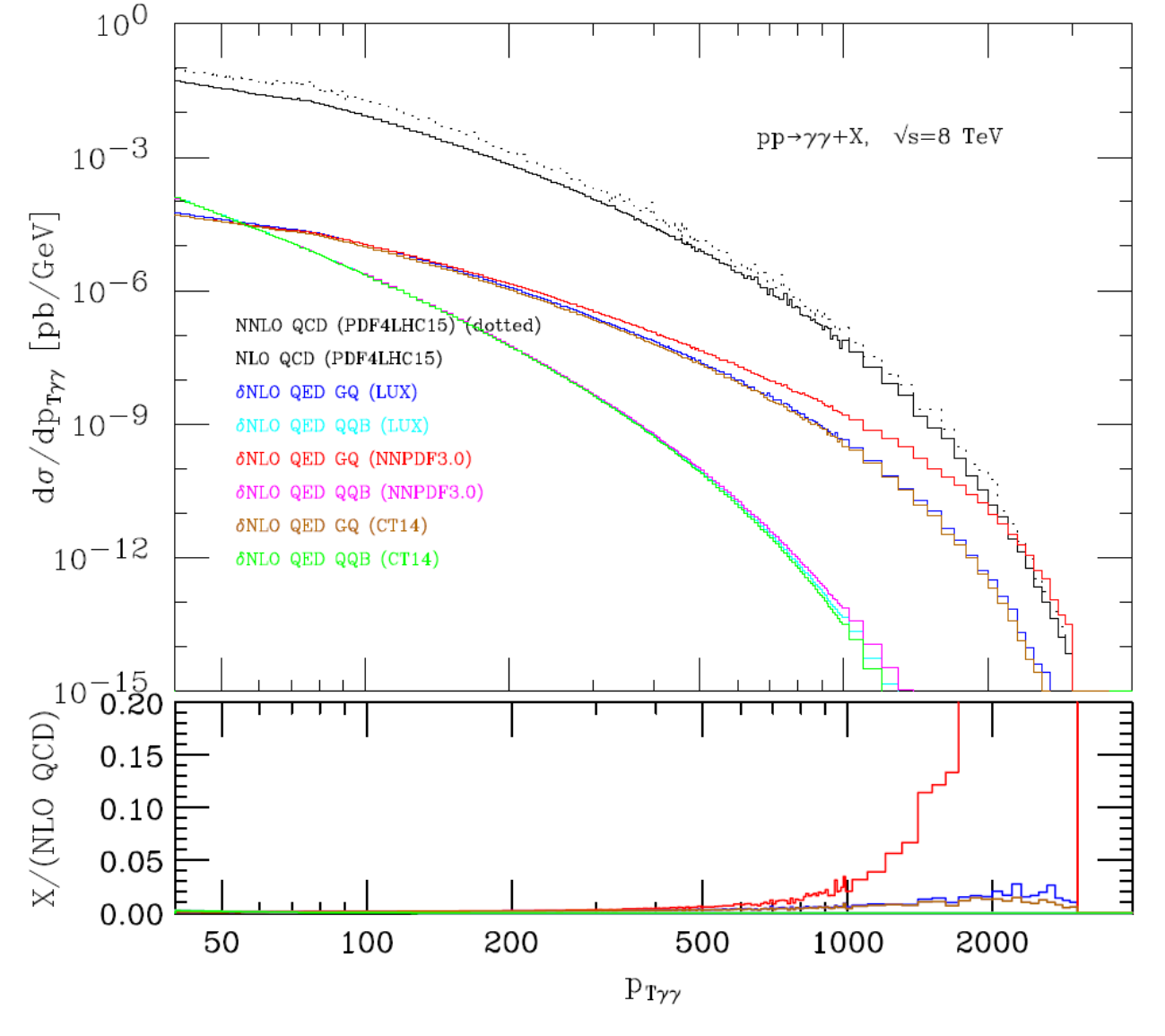} 
\caption{Analysis of the typical distributions for the process $pp \to \gamma \gamma$, including ${\cal O}(\alpha)$ corrections. We present both $M_{\gamma \gamma}$ (left) and $p_T^{\gamma \gamma}$ (right) distributions, showing separately the different production channel (i.e. $q \bar q$ and $q \gamma$) and considering several PDFs sets. The black curves show the NLO (solid) and NNLO QCD (dots) contributions.}
\label{fig:Difotones}
\end{center}
\end{figure} 

After implementing the corresponding experimental cuts, we study the invariant mass and $p_T$ distributions of the diphoton system. The results are shown in Fig. \ref{fig:Difotones}, where we include the NLO QED corrections for each partonic channel (i.e. $q\bar q$ and $q \gamma$, coloured lines), as well as the total NLO and NNLO QCD contributions (black solid and dotted lines). We used \texttt{PDF4LHC} \cite{Rojo:2015acz} for the QCD calculation, and we varied the PDF set of the QED contribution in order to explore the effects of changing the photon PDF. From the plot, we can appreciate that both \texttt{CT14} \cite{Schmidt:2015zda} and \texttt{LUXqed} \cite{Manohar:2016nzj,Manohar:2017eqh} produce similar outputs in the $q \gamma$ channel, but \texttt{NNPDF3.0QED} \cite{Ball:2014uwa} exhibits a completely different behaviour. In particular, the last set strongly enhances the QED corrections in the high-energy region: this effect is still compatible with the behaviour found with \texttt{LUXqed} due to the high uncertainties in the determination of photon distributions. On the other hand, all the distributions almost agree for the $q \bar q$ channel, which shows that quark PDFs are well constrained by the experimental data available.

As a final comment, we would like to emphasize that there are some additional non-trivial features of the higher-order QED corrections. For instance, it is mandatory to properly deal with the EM running coupling, since it could introduce ${\cal O}(10 \, \%)$ deviations in the high-energy distributions. Also, the presence of QED radiation forces the introduction of additional cuts and clustering algorithms, whose phenomenological impact might not be underestimated. A more detailed discussion of this topics can be found in Ref. \cite{INPREP}. 

\section{Conclusions}
\label{sec:conclusions}
In this work, we discussed some features of the computation of higher-order QED corrections. In particular, we focused on the extension of DGLAP equations to deal with the novel lepton and photon distributions, as well as in the computation of the corresponding splitting kernels. We use an Abelianization technique to recover ${\cal O}(\alpha \alphas)$ and ${\cal O}(\alpha^2)$ corrections by making use of the well-known NLO QCD corrections for the AP kernels.

After that, we extended the application of the Abelianization technique to the $q_T$-subtraction method and we obtained the corresponding algorithm to compute NLO QED corrections. We applied this framework to the process $pp\to \gamma \gamma + X$; in particular, we modified the \texttt{2gNNLO} code keeping only the NLO contributions and we verified the consistent cancellation of IR singularities. Moreover, we used this implementation to explore some phenomenological aspects of the process. For instance, we studied the dependence on the PDF set (focusing in the photon distribution), the high-energy behaviour of the QED corrections and the implementation of experimental cuts when QED radiation is included. 

From this analysis, we conclude that QED corrections must be seriously taken into account in the context of high-precision physics. In fact, recent studies for diphoton plus jets \cite{Chiesa:2017gqx} confirm the relevance of the previous assertion and the necessity of a proper understanding of the EW contributions in the high-energy region.

\section*{Acknowledgments}
I would like to thank F. Driencourt-Mangin for carefully reading this article and suggesting modifications. This work has been done in collaboration with D. de Florian and G. Rodrigo (splitting functions), and with L. Cieri and G. Ferrera (diphoton corrections). The research project was partially supported by CONICET, ANPCyT, the Spanish Government, EU ERDF funds (grants FPA2014-53631-C2-1-P and SEV-2014-0398) and Fondazione Cariplo under the grant number 2015-0761.

\end{document}